\documentclass[preprint,authoryear, 12pt]{elsarticle}

\makeatletter
\def\ps@pprintTitle{%
 \let\@oddhead\@empty
 \let\@evenhead\@empty
 \def\@oddfoot{}%
 \let\@evenfoot\@oddfoot}
\makeatother

\usepackage{amsthm,amsmath,amssymb,mathtools}
\usepackage{bbm}
\usepackage{graphicx}
\usepackage{natbib}
\usepackage{geometry}
\usepackage{booktabs, subcaption} 
\usepackage{ragged2e}
\usepackage{hyperref}
\usepackage{xcolor}
\usepackage{soul}
\hypersetup{
    colorlinks=true,
}
\newtheorem{theorem}{Theorem}

\newtheorem{corollary}{Corollary}

\newtheorem{assumption}{Assumption}
\newtheorem{remark}{Remark}

\newcommand{\ind}{\mathbbm{1}}
\newcommand{\Var}{\operatorname{Var}}

\newcommand{\plim}{\overset{p}{\longrightarrow}}

\begin{document}

\begin{frontmatter}

\title{Inference for Batched Adaptive Experiments\tnoteref{t1}}
\tnotetext[t1]{We thank Leif Döring, Taro Fujita, Michael Knaus, Kei Hirano, Jack Porter, David Preinerstorfer, Christoph Rothe, Richard Winter, Kelly W. Zhang for insightful comments and discussions, as well as participants at various seminars and conferences for helpful comments. 
Davud Rostam-Afschar thanks the Deutsche Forschungsgemeinschaft (DFG, German Research Foundation) for financial support through CRC \textit{TRR 266 Accounting for Transparency} (Project-ID 403041268). 
We declare that we have no interests, financial or otherwise, that relate to the research described in this paper.}

\author[addr1]{Jan Kemper}
\ead{jan.kemper@zew.de}

\author[addr2]{Davud Rostam-Afschar\corref{cor1}}
\ead{rostam-afschar@uni-mannheim.de}

\address[addr1]{University of Mannheim, ZEW}
\address[addr2]{University of Mannheim, IZA@LISER, GLO, Labor Hub, NeSt}
\cortext[cor1]{Corresponding author}

\begin{abstract}
The advantages of adaptive experiments have led to their rapid adoption in economics, other fields, and among practitioners. However, adaptive experiments pose challenges for causal inference. This note suggests a BOLS (batched ordinary least squares) test statistic for inference on treatment effects in adaptive experiments. The statistic provides a precision-equalizing aggregation of per-period treatment-control differences under heteroskedasticity. The combined test statistic is a linear combination of heteroskedastic per-period $z$-statistics and can be used to construct asymptotically valid confidence intervals. We provide simulation results comparing rejection rates in the typical case with few treatment periods and few (or many) observations per batch.
\end{abstract}


\end{frontmatter}


\section{Introduction}

Adaptive experiments have become increasingly common because they allow for \emph{earning while learning}. Such designs have been applied, for example, by \citet{Kasy2021}, \citet{Caria2023}, \citet{Offer2021}, \cite{Avivi2021}, \cite{Tabord2023}, \citet{Hoffmann2023}, \citet{Gaul2025}, \citet{Kemper2026}. They combine exploration and exploitation by updating treatment probabilities based on accumulated evidence. However, the dependence of assignment on past outcomes breaks the usual assumptions of random sampling and independent treatment assignment, complicating statistical inference. This is particularly problematic when there is no clear difference between outcomes under different treatments. For example, usual confidence intervals and bootstrap methods may overreject null hypotheses. \cite{Hadad2021} use large-number-of-trials asymptotics to construct generally valid confidence intervals. \citet{Zhangetal2020} note that the number of trials is typically limited, but treatment assignment is adapted after each batch of observations arrives. 

This note extends their argument to the more general and empirically relevant case of heteroskedastic outcomes, deriving the corresponding BOLS (batched ordinary least squares) test statistic and exploring its asymptotic distribution. The heteroskedastic case is relevant because researchers usually design experiments such that not only the outcome means but also their variances may differ by treatment arm. Often the outcome is binary (success/failure), which results in heteroskedasticity by construction.

The failure of standard procedures stems from the behavior of the treatment assignment probabilities, $\pi_t$. Under a zero-margin null hypothesis where treatments perform similarly, adaptive algorithms do not settle on a single optimal arm. Consequently, $\pi_t$ does not concentrate on a deterministic constant but instead converges in distribution to a random variable. Under homoskedasticity, the weighting scheme proposed by \citet{Zhangetal2020} relies on the cancellation of this random limit. However, we show that under heteroskedasticity ($\sigma_1^2 \neq \sigma_0^2$), this cancellation fails. The asymptotic variance of the standard BOLS estimator remains a random variable, causing the estimator to converge to a non-normal limit.

We solve this by proposing a \textit{locally studentized} BOLS statistic. By standardizing the treatment effect by its heteroskedastic standard error within each batch \emph{before} aggregation, we mechanically force the predictable variance contribution of each batch to be equal. This eliminates the random limits of the assignment probabilities from the asymptotic distribution, restoring asymptotic normality and ensuring valid inference. The results are in line with \cite{Hirano2025}, who show that any limit distribution generated by joint choices of adaptive assignment rules and statistics can be represented within a unified Gaussian limit experiment \citep[see also][]{Bibaut2025}. However, \cite{Hirano2025} do not study BOLS estimators, derive precision-equalizing weights, or analyze finite-sample performance in adaptive bandit designs. We fill this gap by deriving a heteroskedastic extension of the BOLS test statistic with precision-equalizing weights. Our approach yields a simple, implementable inference procedure and shows when heteroskedasticity matters in adaptive experiments.

\section{Treatment Effects in Adaptive Experiments}
\label{Treatment Effects in Adaptive Experiments}
\subsection{Target}
The main target of this analysis is to show that a test statistic $Z_{het}$, aggregated across batches, is asymptotically standard normal in batched adaptive experiments under heteroskedasticity. In applied settings, the most relevant application of this statistic is to test whether the treatment effect (e.g. treatment vs control arm) under homogeneous treatment effects is statistically different from zero across batches.
\subsection{Setting and outcomes}
Consider an experiment where in batches $t=1,\ldots,T$, participants $i=1,\ldots,n_t$ are assigned to treatment arm $a\in\{0,1\}$. Their potential outcomes are
\begin{equation}\label{eq:potentialoutcomes}
    Y_{i,t}(a)=\mu_{a,t}+\varepsilon_{i,t}(a).
\end{equation}

Let $\varepsilon_{i,t}(a)$ denote an arm-specific error term (heteroskedasticity) and $\Delta_t=\mu_{1,t}-\mu_{0,t}$ denote the batch-specific treatment effect. We test the joint null $H_0:\Delta_t=\delta_{0,t}$ for all $t$. A relevant special case is that of homogeneous treatment effects $\Delta_t=\Delta$, which could be tested against $\delta_{0,t}=0$ for all $t$.

\subsection{Treatment assignment}

Let $H_{t-1} = \{(Y_{i,s}, A_{i,s}): i=1,\ldots,n_s;\; s=1,\ldots,t-1\}$ denote the history of outcomes and treatment assignments from batches $1,\ldots,t-1$. The adaptive assignment mechanism is characterized as follows:

\begin{itemize}
    \item[(i)] The treatment probability for batch $t$ is $\pi_t = P(A_{i,t} = 1 \mid H_{t-1})$, which is $H_{t-1}$-measurable.
    \item[(ii)] Within batch $t$, treatment assignments are conditionally i.i.d.: $A_{1,t}, \ldots, A_{n_t,t} \mid H_{t-1} \stackrel{i.i.d.}{\sim} \text{Bernoulli}(\pi_t)$.
    \item[(iii)] Given $(H_{t-1}, A_{i,t})$, outcomes $Y_{i,t}$ are independent across individuals within batch $t$.
\end{itemize}

\subsection{Treatment effects}

In period $t$, there are $N_{1,t}=\sum_{i=1}^{n_t}A_{i,t}$ treated and $N_{0,t}=n_t-N_{1,t}$ control units. We distinguish the algorithmic assignment probability $\pi_t=P(A_{i,t}=1\mid H_{t-1})$ from the realized treatment share $\hat\pi_t=N_{1,t}/n_t$.

Let the per-period difference in sample means be
\begin{equation}\label{eq:samplemeans}
    \hat{\Delta}_t = \frac{1}{N_{1,t}} \sum_{i=1}^{n_t} A_{i,t} Y_{i,t} - \frac{1}{N_{0,t}} \sum_{i=1}^{n_t} (1 - A_{i,t}) Y_{i,t}=\bar Y_{1,t}-\bar Y_{0,t}.
\end{equation}
Let $\mathcal{G}_t = H_{t-1} \vee \sigma(A_{1,t},\ldots,A_{n_t,t})$ denote the information available after the batch-$t$ assignments are realized but before the batch-$t$ outcomes are known. Conditional on $\mathcal{G}_t$, the realized treatment counts $N_{1,t}$ and $N_{0,t}$ are fixed. Under conditional independence and exogeneity to period-$t$ potential outcome shocks,
\begin{equation}\label{eq:variancesbyarms}
    \Var(\bar Y_{1,t}\mid \mathcal{G}_t)=\frac{\sigma_{1,t}^2}{N_{1,t}}, \qquad
    \Var(\bar Y_{0,t}\mid \mathcal{G}_t)=\frac{\sigma_{0,t}^2}{N_{0,t}}.
\end{equation}

Hence, the conditional variance of the period difference is
\begin{align}
    v_t \;\equiv\; \Var(\hat\Delta_t - \Delta_t \mid \mathcal{G}_t) 
    &= \frac{\sigma_{1,t}^2}{N_{1,t}}+\frac{\sigma_{0,t}^2}{N_{0,t}} = \frac{1}{n_t}\!\left(\frac{\sigma_{1,t}^2}{\hat\pi_t}+\frac{\sigma_{0,t}^2}{1-\hat\pi_t}\right). \label{eq:vt}
\end{align}

\section{Inference}

\subsection{The batchwise and the BOLS test statistics}

In adaptive experiments, the assignment probability $\pi_t$ is random because it depends on the realized history. Consequently, the realized allocation shares $\hat\pi_t$ and the variances of batchwise estimators are random. This may result in asymptotic non-normality for non-studentized estimators. Intuitively, if outcomes under two treatments are hard to distinguish, either treatment might get assigned more observations in repeated samples, and consequently the assignment probability does not concentrate. \cite{Zhangetal2020} show that the selection probability is fixed when conditioning on the history up to a given batch, and that the homoskedastic batchwise OLS, scaled by the selection probability, is asymptotically normal. 
Under heteroskedasticity, this scaling no longer removes the random assignment probabilities from the limiting variance. We therefore base inference not on an estimator but on a z-statistic that accounts for heteroskedasticity.

The key idea is to studentize within each batch \emph{before} aggregation. 
For each batch $t$, define the $z$-score
\begin{equation}\label{eq:zt}
    z_{t,\mathrm{het}} = \frac{\hat\Delta_t - \delta_{0,t}}{\sqrt{v_t}},
\end{equation}

where $\delta_{0,t}$ is the null value of the  treatment effect in batch $t$. The variance $v_t$ is random, but the conditional variance of each score $\Var(z_{t,\mathrm{het}}\mid \mathcal{G}_t) = \Var(\hat\Delta_t-\Delta_t\mid\mathcal{G}_t)/v_t = 1$ is not. This is the crucial property: local studentization eliminates the random assignment probabilities from the variance. Therefore, $\sum_{t=1}^T\Var(z_{t,\mathrm{het}}\mid \mathcal{G}_t) =T.$

Our objective is to derive a test statistic aggregated across batches $Z_{\mathrm{het}}$ that is asymptotically $\mathcal{N}(0,1)$ under heteroskedasticity. The BOLS test statistic is the sum of batchwise $z$-scores, normalized by the \emph{deterministic} factor $1/\sqrt{T}$:
\begin{equation}\label{eq:Zhet}
    Z_{\mathrm{het}} = \frac{1}{\sqrt{T}}\sum_{t=1}^T z_{t,\mathrm{het}}= \frac{1}{\sqrt{T}}\sum_{t=1}^T \frac{\hat\Delta_t - \delta_{0,t}}{\sqrt{v_t}}.
\end{equation}
The martingale-difference property implies that scores from different batches are uncorrelated, such that $Z_{\mathrm{het}}$ has unit variance $\Var(Z_{\mathrm{het}})=T^{-1}\sum_{t=1}^T E\big[\Var(z_{t,\mathrm{het}}\mid \mathcal{G}_t)\big]=1$. 

The statistic serves two purposes. It can be used to test the batchwise null $H_0:\Delta_t=\delta_{0,t}$ for all $t=1,\ldots,T$ (Section~\ref{sec:asymptotic}). Under the common-effect restriction $\Delta_t=\Delta$ for all $t$, this null is $H_0:\Delta=\delta_0$.
Additionally, we derive a point estimate and a confidence interval for $\Delta$ (Section~\ref{sec:confidence interval}).

\subsection{Asymptotic theory}\label{sec:asymptotic}

\begin{assumption}[Conditional Independence and Exogeneity]\label{ass:indep}
Within batch $t$, conditional on $H_{t-1}$, the potential outcomes $\{Y_{i,t}(0),Y_{i,t}(1)\}_{i=1}^{n_t}$ are independent across individuals and independent of the randomization used to generate the treatment assignments in batch $t$.
\end{assumption}

\begin{assumption}[Moment Conditions]\label{ass:moments}
For all $t$ and $a \in \{0,1\}$: (i) $E[\varepsilon_{i,t}(a) \mid H_{t-1}] = 0$ and $E[\varepsilon_{i,t}(a)^2 \mid H_{t-1}] = \sigma_{a,t}^2$; (ii) there exists some $\delta > 0$ such that $E[|\varepsilon_{i,t}(a)|^{2+\delta} \mid H_{t-1}] < M < \infty$ almost surely; and (iii) the variances are nondegenerate and uniformly bounded, $0<\underline\sigma^2\leq\sigma_{a,t}^2\leq\bar\sigma^2<\infty$.
\end{assumption}

\begin{assumption}[Clipping Constraint]\label{ass:clip}
There exists a constant $\kappa \in (0, 0.5]$ such that $P(\pi_t \in[\kappa, 1-\kappa]) \to 1$ as $n_t \to \infty$ for all $t$.
\end{assumption}

\begin{assumption}[Consistent Variance Estimation]\label{ass:var}
The variance estimators satisfy $\hat\sigma_{a,t}^2 \plim \sigma_{a,t}^2$ for all $t$ and $a \in \{0,1\}$.
\end{assumption}

Assumptions \ref{ass:indep}--\ref{ass:clip} correspond to Conditions 1, 3, and the clipping constraint (Definition 1) in \citet{Zhangetal2020}, extended to allow $\sigma_{1,t}^2 \neq \sigma_{0,t}^2$.

\begin{theorem}[Asymptotic Normality of the Heteroskedastic BOLS Statistic]\label{thm:main}
Under Assumptions~\ref{ass:indep}--\ref{ass:clip}, consider the batchwise null
hypothesis $H_0:\Delta_t=\delta_{0,t}$ for all $t=1,\ldots,T$, with null values
$\delta_{0,1},\ldots,\delta_{0,T}$. Define the infeasible period scores and
combined statistic as
\begin{equation}\label{eq:zhet}
    z_{t,\mathrm{het}} = \frac{\hat\Delta_t-\delta_{0,t}}{\sqrt{v_t}},\qquad
    Z_{\mathrm{het}} = \frac{1}{\sqrt{T}}\sum_{t=1}^T z_{t,\mathrm{het}}.
\end{equation}
Then, under $H_0$, for fixed $T$ and $\min_t n_t \to \infty$,
\begin{equation}\label{eq:zhetnormal}
    Z_{\mathrm{het}} \;\overset{d}{\longrightarrow}\; \mathcal{N}(0,1).
\end{equation}

If, in addition, Assumption~\ref{ass:var} holds and
$\hat v_t = \hat\sigma_{1,t}^2/N_{1,t} + \hat\sigma_{0,t}^2/N_{0,t}$, then the
feasible statistic
\begin{equation}\label{eq:zhetfeasible}
    \hat Z_{\mathrm{het}} = \frac{1}{\sqrt{T}}\sum_{t=1}^T
    \frac{\hat\Delta_t-\delta_{0,t}}{\sqrt{\hat v_t}}
\end{equation}
also satisfies $\hat Z_{\mathrm{het}} \overset{d}{\longrightarrow} \mathcal{N}(0,1)$
under $H_0$. If, in addition, treatment effects are assumed common across batches, \(\Delta_t=\Delta\) for all \(t\), then the same statistic tests $H_0:\Delta=\delta_0$. The proof is provided in~\ref{app:proof}.
\end{theorem}

\noindent\textbf{Intuition of the proof.} We studentize the estimates \textit{locally}. By dividing the batchwise effect by its heteroskedastic standard error \emph{before} aggregation across batches, we normalize each batch contribution to have predictable variance $1/T$ in the final statistic. The proof treats the adaptively chosen assignment probabilities as random but predictable quantities and applies a triangular-array martingale central limit theorem directly to the individual-level observations across all batches.

\subsection{From BOLS test statistic to confidence intervals}
\label{sec:confidence interval}

Inverting the feasible BOLS test statistic of Theorem~\ref{thm:main} yields a
point estimate and a confidence interval for a \textit{weighted aggregate
effect}. This is a weighted average of the batch-specific effects $\Delta_t$ with weights
proportional to their inverse standard errors. Let
$w_t = 1/\sqrt{v_t}$ denote the inverse standard error of
$\Delta_t$, let $S=\sum_{s=1}^T w_s$, and define the weights
\begin{equation}\label{eq:omega}
    \omega_t = \frac{w_t}{S}, \qquad \omega_t > 0, \qquad
    \sum_{t=1}^T \omega_t = 1 .
\end{equation}
The weighted aggregate effect and its estimator, which we call the
\textit{heteroskedastic BOLS estimator}, are
\begin{equation}\label{eq:theta}
    \theta = \sum_{t=1}^T \omega_t\,\Delta_t ,
    \qquad
    \hat\theta = \sum_{t=1}^T \hat\omega_t\,\hat\Delta_t .
\end{equation}

Without further restrictions, $\theta$ is a weighted average of the
batchwise $\Delta_t$. Since the $\Delta_t$ can differ and the weights
$\omega_t$ depend on the realized allocations, $\theta$ is random and
varies across realizations of the experiment. The restriction to a
homogeneous effect $\Delta_t=\Delta$ for all $t$ ensures that the target is
a fixed parameter. Under this restriction $\theta$ equals $\Delta$
regardless of the weights, and the heteroskedastic BOLS estimator
$\hat\theta$ consistently estimates $\Delta$. The test $H_0:\Delta=\delta_0$ and the corresponding confidence interval refer to a single, well-defined parameter in this case. The following further results are all derived under the \textit{homogeneous treatment effect assumption} ($\Delta_t=\Delta$). Remark~\ref{rem:heterogeneous} discusses the implications for $\hat{\theta}$ when we lift the homogeneous treatment effect assumption.

Evaluating the feasible statistic of Theorem~\ref{thm:main} at
$\delta_{0,t}=\Delta_t$ and writing $\hat q_T=\sqrt{T}/\hat S$, it follows
from~\eqref{eq:theta} that
\begin{equation}\label{estimator_zstat}
    \hat Z_{\mathrm{het}}
    = \frac{1}{\sqrt{T}}\sum_{t=1}^T \frac{\hat\Delta_t-\Delta_t}{\sqrt{\hat v_t}}
    = \frac{1}{\sqrt{T}}\sum_{t=1}^T \hat w_t(\hat\Delta_t-\Delta_t)
    = \frac{\hat\theta-\theta}{\hat q_T}.
\end{equation}

\begin{corollary}[Asymptotically Valid Confidence Interval]\label{cor:ci}
Suppose Assumptions~\ref{ass:indep}--\ref{ass:var} hold and $\Delta_t=\Delta$
for all $t$. Then
\begin{equation}\label{eq:ci}
    \mathrm{CI}_{1-\alpha}=\hat\theta \pm z_{1-\alpha/2}\,\hat q_T
\end{equation}
satisfies $P\bigl(\Delta\in\mathrm{CI}_{1-\alpha}\bigr)\to 1-\alpha$, where
$z_{1-\alpha/2}$ is the $(1-\alpha/2)$ quantile of the standard normal
distribution.
\end{corollary}

Coverage follows by inversion from~\eqref{estimator_zstat} and Theorem~\ref{thm:main}, since
$(\hat\theta-\theta)/\hat q_T\overset{d}{\longrightarrow}\mathcal N(0,1)$.

\subsection{Properties of the heteroskedastic BOLS estimator}
\label{sec:Properties of the BOLS estimator}

The heteroskedastic BOLS estimator is a weighted average of the batchwise estimates $\hat\Delta_t$
with data-dependent weights $\hat\omega_t$. Its estimation error is
\begin{equation}\label{eq:error}
    \hat\theta-\theta=\sum_{t=1}^T\hat\omega_t\,(\hat\Delta_t-\Delta_t),
\end{equation}
which follows from~\eqref{eq:theta}. 

\begin{remark}[Bias]\label{rem:Bias}
Each batchwise estimator is conditionally unbiased,
$E[\hat\Delta_t-\Delta_t\mid\mathcal{G}_t]=0$. The heteroskedastic BOLS estimator
nevertheless need not be unbiased, $E[\hat\theta-\theta]\neq 0$ in
general, because future weights depend on past outcomes and thus the weights in~\eqref{eq:error} are correlated with the estimation errors.
\end{remark}

\begin{remark}[Consistency]\label{rem:Consistency}
Despite its finite-sample bias, $\hat\theta$ is consistent under homogeneous effects. Since
$\hat\omega_t\geq 0$ and $\sum_{t=1}^T\hat\omega_t=1$ by~\eqref{eq:omega},
\eqref{eq:error} gives
\begin{equation}\label{consistencyproof}
    |\hat\theta-\theta|
    \leq \sum_{t=1}^T\hat\omega_t\,|\hat\Delta_t-\Delta_t|
    \leq \max_{t}|\hat\Delta_t-\Delta_t|.
\end{equation}
The first inequality comes from the triangle inequality that implies that the absolute aggregate error does not exceed the sum of the individual absolute errors; the second holds because a
weighted average cannot exceed its largest element. These inequalities hold in finite samples, regardless of how the adaptive weights defined in \eqref{eq:omega} are determined. Hence, for fixed $T$ and
$n=\min_t n_t\to\infty$,
\begin{equation}\label{consistency}
    \hat\theta-\theta=O_p(n^{-1/2}),
    \qquad \hat\theta\xrightarrow{p}\Delta.
\end{equation}

\end{remark}

\begin{remark}[Distribution of $\hat\theta$]\label{rem:asyvar}
In general, $\hat\theta$ is not asymptotically normal, because the adaptive
weights $\hat\omega_t$ remain random in large samples. This does not affect the
confidence interval in Corollary~\ref{cor:ci}, which relies only on the
asymptotic normality of $\hat Z_{\mathrm{het}}=(\hat\theta-\theta)/\hat q_T$. If
the weights are fixed, $\hat\theta-\theta=\sum_t\omega_t(\hat\Delta_t-\Delta_t)$
is a fixed linear combination of asymptotically independent normal batch
estimates and hence asymptotically normal, with variance $\sum_t\omega_t^2v_t=T/S^2$. When the treatment effect is large relative to
the outcome noise, the algorithm quickly identifies the better arm, the
assignment probabilities settle at a deterministic value, and $\hat\theta$ is
approximately normal.
\end{remark}

\begin{remark}[Heterogeneous treatment effects]\label{rem:heterogeneous}
Under heterogeneous treatment effects, $\hat\theta$ does not converge to a
fixed parameter, because under adaptive assignment the weights remain
random in large samples, so that $\hat\theta$ converges to different linear
combinations of the $\Delta_t$ across realizations of the experiment.
Nonetheless, Theorem~\ref{thm:main} remains valid, and a confidence
interval can be derived for the linear combination given by the realized
weights. This interval is, however, difficult to interpret, since its
target is not a fixed parameter but depends on the history of the
experiment.
\end{remark}

\subsection{Feasible implementation}\label{sec:feasible}

The heteroskedastic BOLS estimator~\eqref{eq:theta} and the confidence interval in
Corollary~\ref{cor:ci} require only the batchwise estimates $\hat\Delta_t$ and the
variance estimates
\begin{equation}\label{estimatedvariance}
    \hat v_t = \frac{\hat\sigma_{1,t}^2}{N_{1,t}} + \frac{\hat\sigma_{0,t}^2}{N_{0,t}},
\end{equation}
where $\hat\sigma_{a,t}^2$ is the sample variance of the outcomes in arm $a$ of
batch $t$. Table~\ref{tab:weights}
shows the resulting weights for the non-adaptive and adaptive cases under homoskedasticity and heteroskedasticity.

\begin{remark}[Variance Estimation and Weights]\label{rem:hc}
(i) In small samples, finite-sample adjusted variance estimators (e.g.,
HC2/HC3) may be preferable.
(ii) If the variances are constant across batches,
$\sigma_{a,t}^2=\sigma_a^2$, arm-specific variance estimators pooled across
batches also satisfy Assumption~\ref{ass:var} and are more precise
\citep[cf.\ Corollary~4 in][]{Zhangetal2020}. Pooling should use deviations from
the batch-specific arm means, since $\mu_{a,t}$ may vary across batches.
(iii) Under homoskedasticity ($\sigma_{1,t}^2=\sigma_{0,t}^2=\sigma^2$) and
constant batch size, the weights reduce to
$\omega_t = \sqrt{N_{1,t} N_{0,t}}/\sum_{s=1}^T \sqrt{N_{1,s} N_{0,s}}$, so
batches with more balanced realized sizes receive larger weight. With a pooled
variance estimate, the homoskedastic $\hat\sigma^2$ cancels and the feasible weights $\hat\omega_t$
equal this expression exactly.
(iv) The extension to $K$ treatment arms and contextual settings can be constructed following the same arguments applied in this analysis.
\end{remark}

\section{Monte Carlo Simulations}

We compare three test statistics: the heteroskedasticity-robust OLS statistic, the BOLS statistic derived under homoskedasticity, and our heteroskedasticity-robust BOLS statistic. We report \emph{rejection rates}, i.e., the proportion of simulated samples in which the true null hypothesis is rejected. Under the zero-margin null $H_{0}:\,\Delta = 0$, the algorithm struggles to identify the optimal arm, so that treatment probabilities $\pi_t$ remain highly variable. The rejection rate measures the empirical size of the test (nominal 5\%). Data are generated using two common adaptive sampling algorithms, $\varepsilon$-Greedy and Bernoulli Thompson Sampling in two-arm settings.

\begin{figure}[htbp]
  \centering
  \begin{minipage}[b]{0.45\textwidth}
    \centering
    \includegraphics[width=\textwidth]{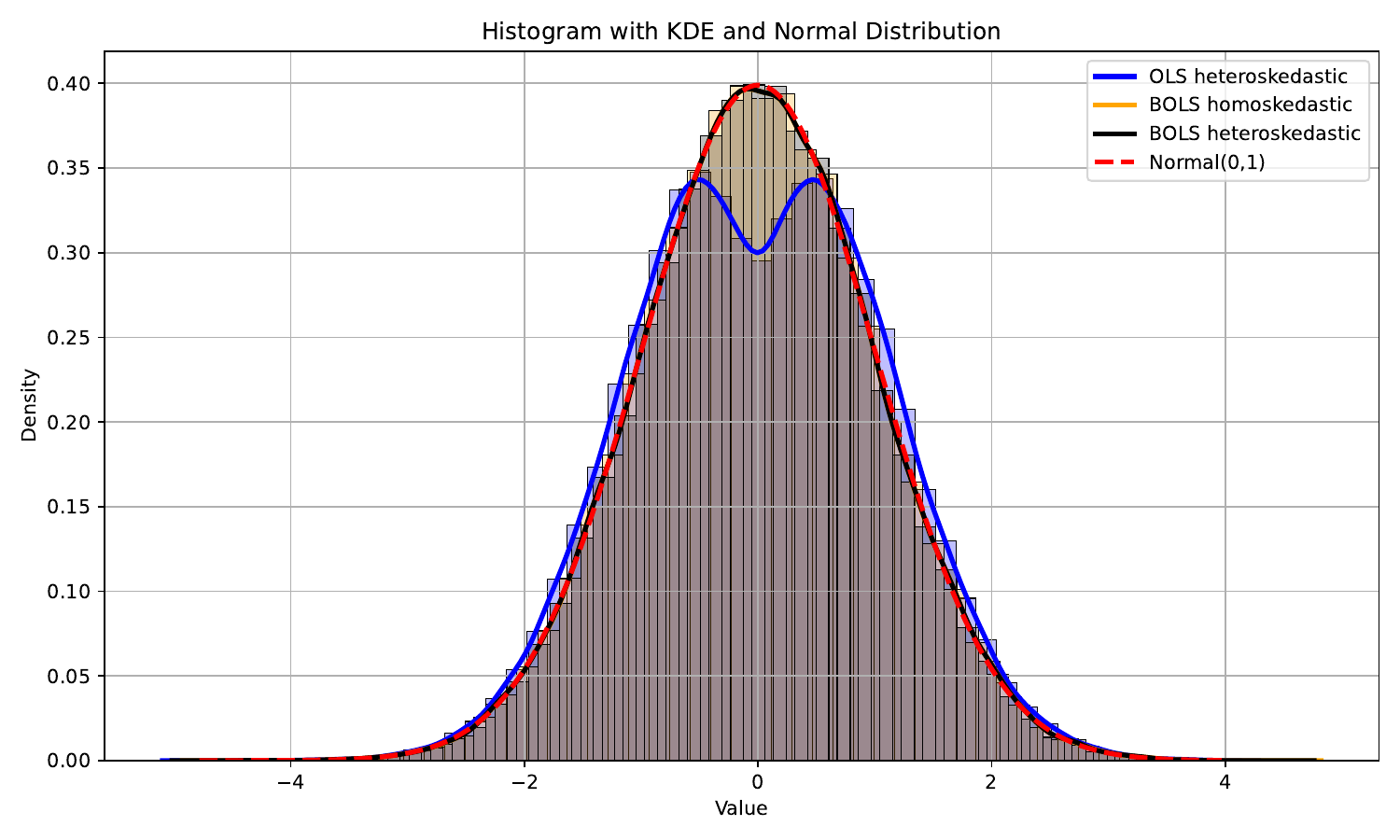}
    \subcaption{$\varepsilon$-Greedy, outcomes homoskedastic across arms}
  \end{minipage}
  \hfill
  \begin{minipage}[b]{0.45\textwidth}
    \centering
    \includegraphics[width=\textwidth]{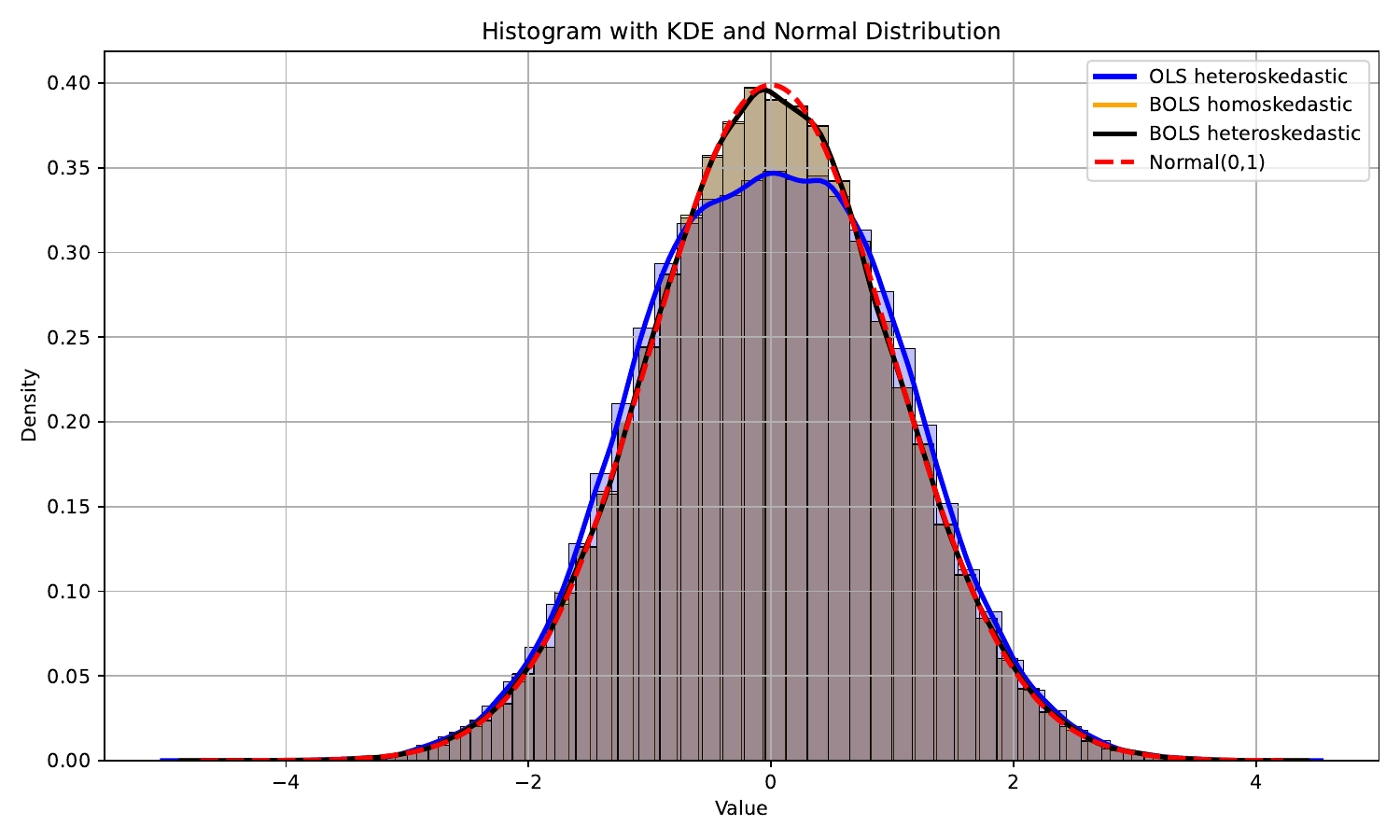}
    \subcaption{Thompson Sampling, homoskedastic across arms}
  \end{minipage}
\\
 \begin{minipage}[b]{0.45\textwidth}
    \centering
    \includegraphics[width=\textwidth]{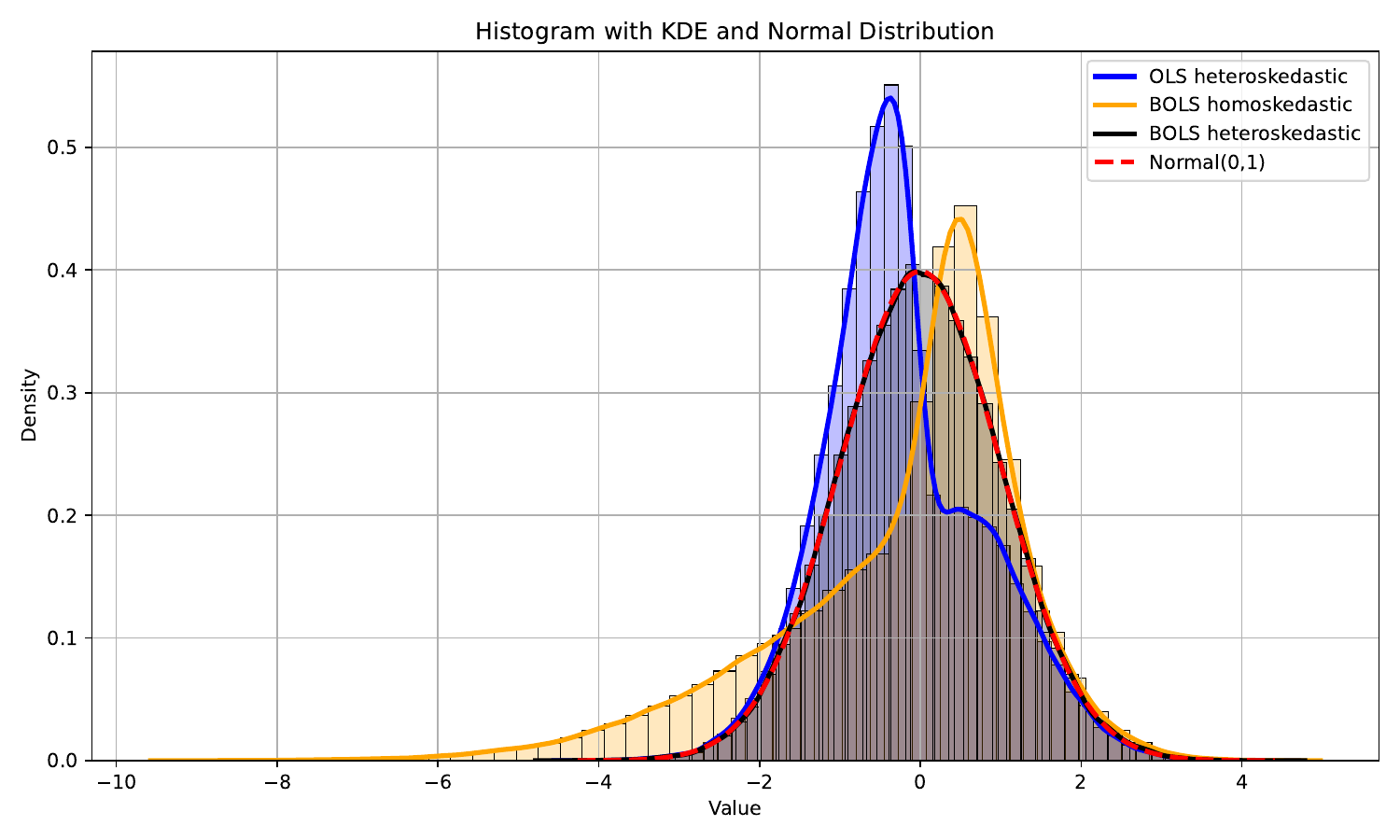}
    \subcaption{$\varepsilon$-Greedy, outcomes heteroskedastic across arms}
  \end{minipage}
  \hfill
  \begin{minipage}[b]{0.45\textwidth}
    \centering
    \includegraphics[width=\textwidth]{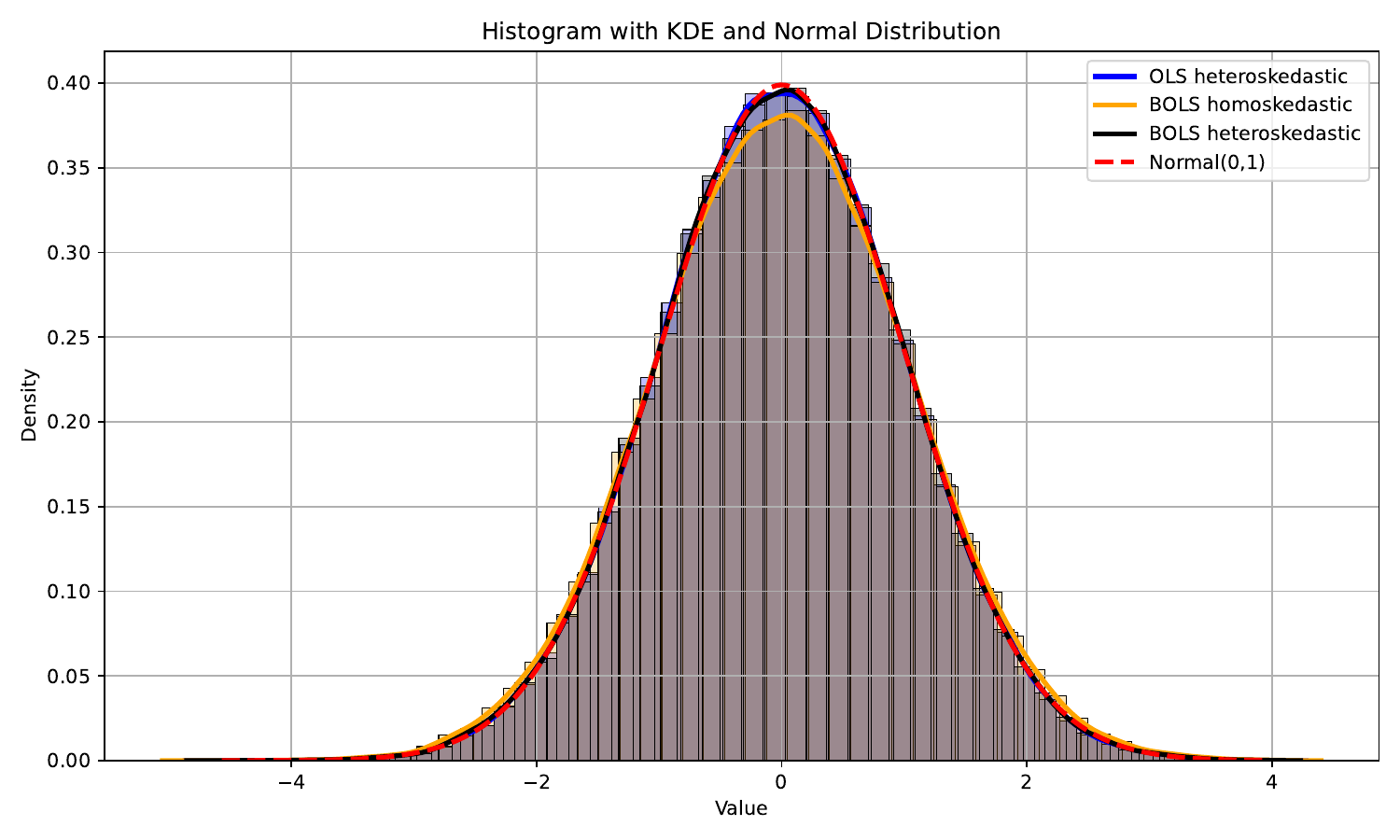}
    \subcaption{Thompson Sampling, heteroskedastic across arms}
  \end{minipage}
  
  \scriptsize
     \begin{justify}
       \textit{Notes:} The figure shows the results of Monte Carlo simulations with 100,000 repetitions. Panels a) and c) show the distribution of the heteroskedasticity-robust OLS test statistic, the homoskedastic BOLS test statistic, and the heteroskedasticity-robust BOLS test statistic for data generated from an $\varepsilon$-Greedy experiment. The batch size is 500, the number of batches 25, the experiment consists of two arms, each with an expected value of 1. In panel a) the standard deviation is 1 for arm one and 1 for arm two (homoskedasticity). In panel c) the standard deviation for arm one is 4 and for arm two is 1, everything else remains equal. The red dotted line indicates the density of the standard normal distribution. Panel b) shows data generated from a Bernoulli Thompson algorithm. The batch size is 500, the number of batches 25, the experiment consists of two arms with an expected value of 0.5 and 0.5 (homoskedasticity). In panel d) the experiment consists of two arms with an expected value of 0.7 and 0.4 (heteroskedasticity). 
          \end{justify}
    \caption{Simulation I: Non-normality of OLS and homoskedastic BOLS}
    \label{fig:sim1}
\end{figure}

\begin{figure}[h!]
    \centering
    \begin{subfigure}[b]{0.32\textwidth}
        \centering
        \includegraphics[width=\textwidth]{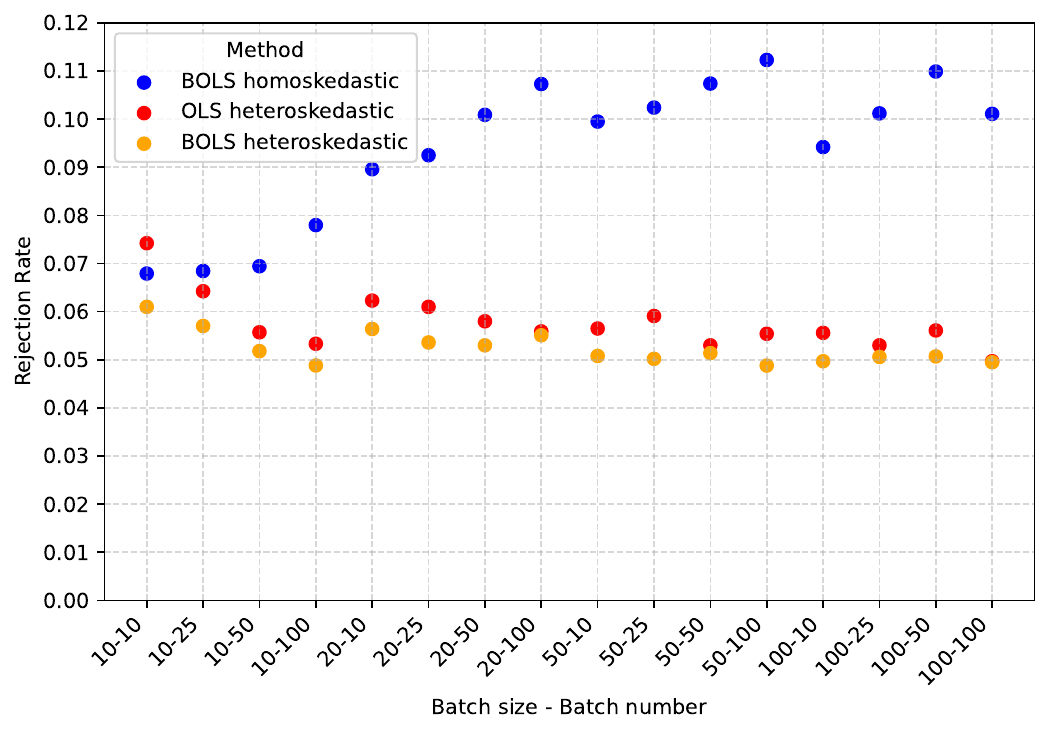}
        \caption{$\Delta = 0$ }
        \label{fig:plot1a}
    \end{subfigure}
    \begin{subfigure}[b]{0.32\textwidth}
        \centering
        \includegraphics[width=\textwidth]{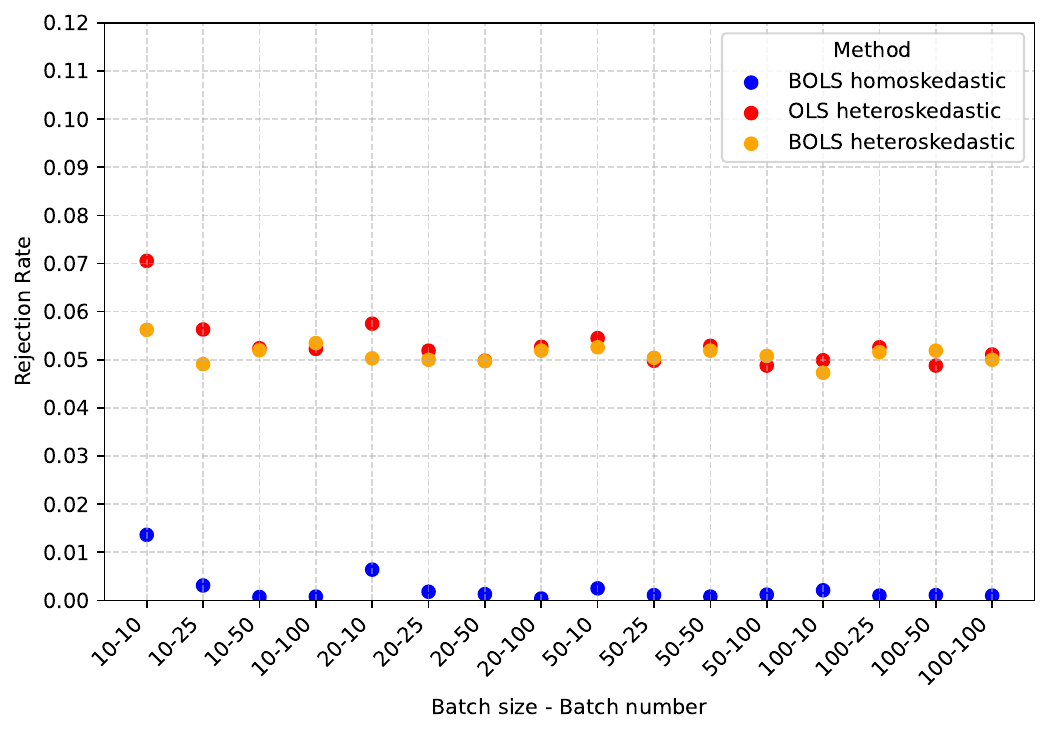}
        \caption{$\Delta = 1$}
        \label{fig:plot1b}
    \end{subfigure}
    \begin{subfigure}[b]{0.32\textwidth}
        \centering
        \includegraphics[width=\textwidth]{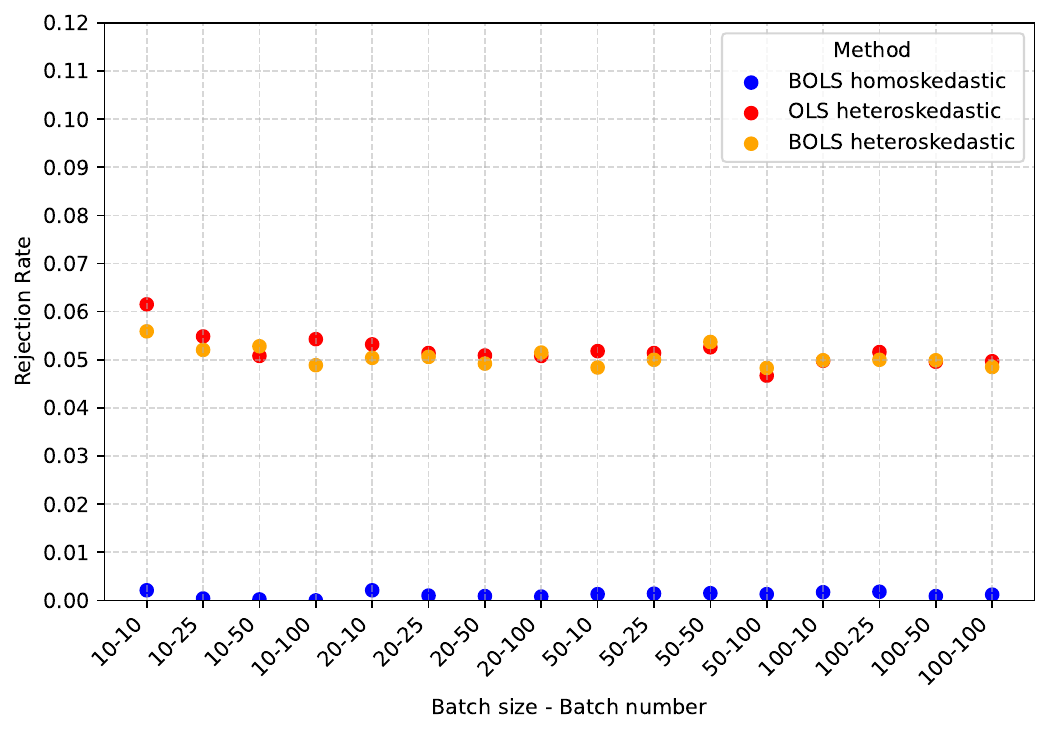}
        \caption{$\Delta = 2$}
        \label{fig:plot1c}
    \end{subfigure}
    \scriptsize
     \begin{justify}
       \textit{Notes:} This figure shows Monte Carlo simulation results for the $\varepsilon$-Greedy algorithm. Combinations of batch size and number of batches increase along the horizontal axis. Rejection rates are denoted on the vertical axis. The parameter $\Delta$ indicates the difference between the true expected values of both arms. Each circle shows the average rejection rate for the given test statistic which is indicated by color. For each batch size / number of batches combinations 10,000 repetitions were executed. In panel (a) the expected value $\mu_1$ for arm $1$ is $1$ and $\mu_2$ for arm $2$ is also $1$. The standard deviation in all panels for arm $1$ is $\sigma_1 = 2$ and for arm $2$ is $\sigma_2 =1$. In panel (b) the expected value is $\mu_1 = 2$ and $\mu_2 = 1$. For panel (c) it is  $\mu_1 = 3$ and $\mu_2 = 1$. In the batch size 10-10 scenario, on average, no test statistic could be calculated for one percent of the draws because one of the arms was never played.
          \end{justify}
    \caption{$\varepsilon$-Greedy}
    \label{fig:sim2}
\end{figure}

\paragraph{Simulation I: Non-normality of test statistics based on OLS and homoskedastic BOLS}

Figure~\ref{fig:sim1} shows the empirical distributions of the test statistics with 25 batches each with a size of 500 observations. For $\varepsilon$-Greedy (left panels), both arms have Gaussian outcomes (think of log incomes) with mean~1, with variances $(1^2, 1^2)$ in panel a) and $(4^2,1^2)$ in panel c). The exploration rate is $\varepsilon=0.2$. Each design is repeated 100{,}000 times. Consistent with \citet{Zhangetal2020}, panel a) shows that the test statistic from OLS under a zero-margin is non-normally distributed and homoskedastic BOLS recovers its normality. But in the same setting under heteroskedasticity (panel c), both OLS and the homoskedastic BOLS statistics deviate markedly from normality in the zero-margin case. The homoskedastic BOLS statistic severely overrejects (17\% instead of 5\%). OLS yields approximately correct rejection rates but exhibits non-normal behavior. In contrast, our heteroskedasticity-robust BOLS statistic closely matches the standard normal distribution and delivers correct 5\% rejection rates in both cases.

For Thompson Sampling (right panels), we set Bernoulli success probabilities to $(p_1,p_2)=(0.5,0.5)$ in panel b) and to $(p_1,p_2)=(0.7,0.4)$ in panel d). Panel b) confirms that the test statistic based on OLS is non-normally distributed and the homoskedastic BOLS test statistic fixes the problem. However, under mechanical heteroskedasticity (panel d), the homoskedastic BOLS statistic overrejects ($\approx6\%$) because it ignores heteroskedasticity. As the success probabilities differ substantially, both OLS and our heteroskedasticity-robust BOLS statistic closely match the standard normal distribution and deliver correct 5\% rejection rates.

\begin{figure}[h!]
    \centering
    \begin{subfigure}[b]{0.32\textwidth}
        \centering
        \includegraphics[width=\textwidth]{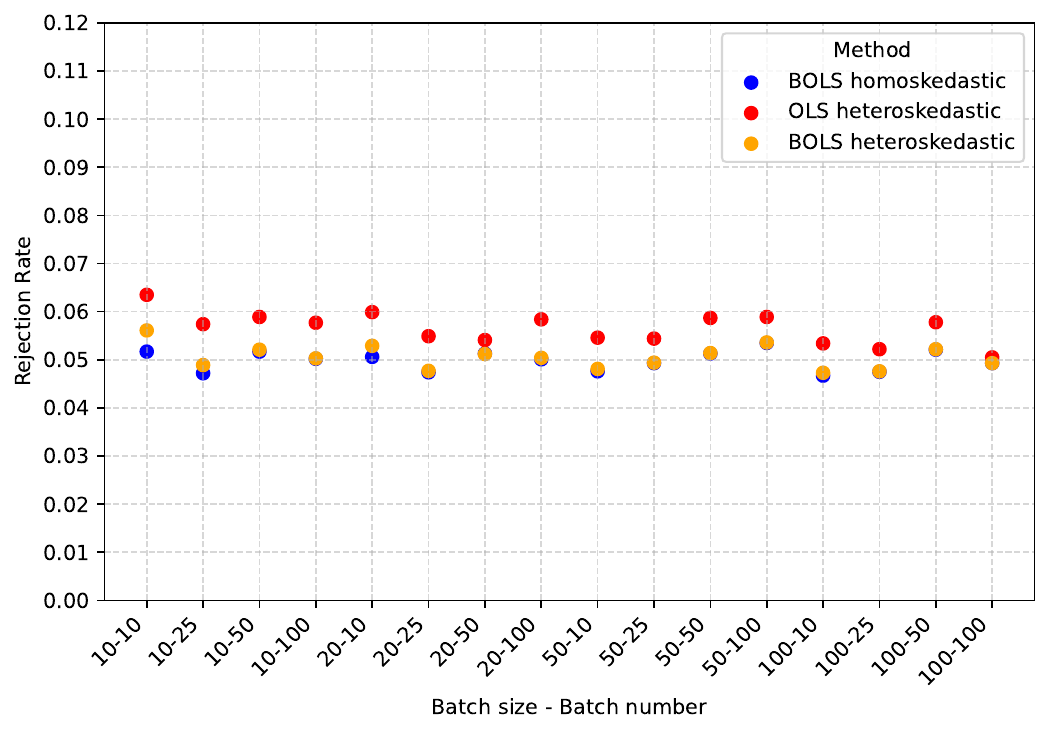}
        \caption{$\Delta = 0$ }
        \label{fig:plot2a}
    \end{subfigure}
    \begin{subfigure}[b]{0.32\textwidth}
        \centering
        \includegraphics[width=\textwidth]{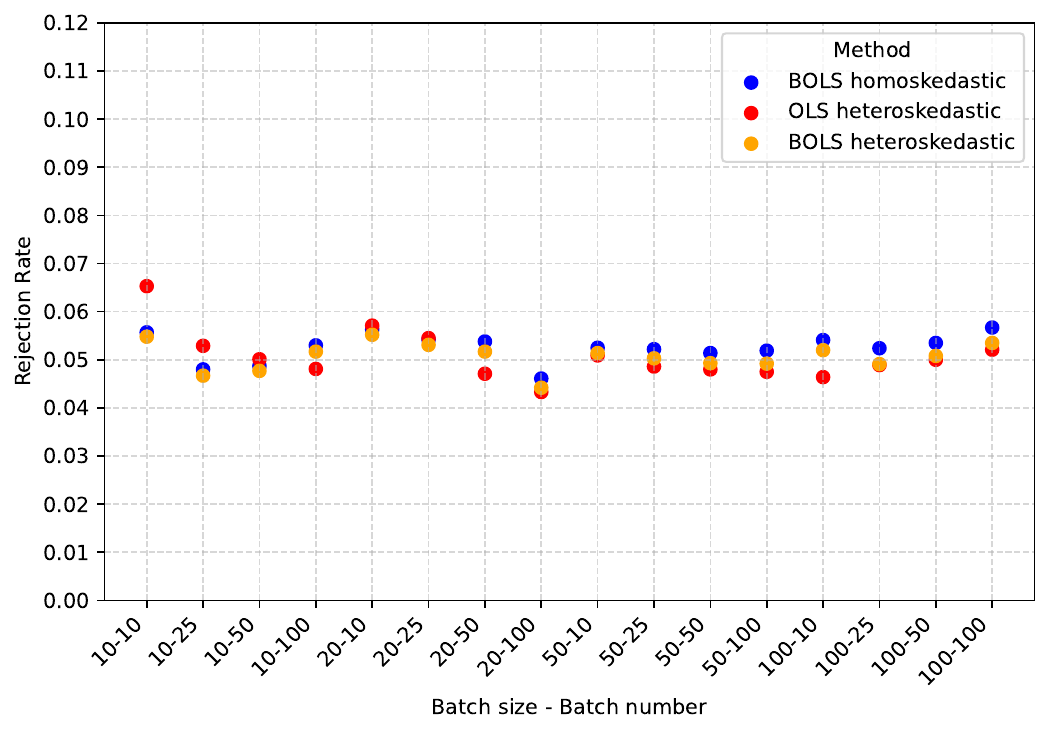}
        \caption{$\Delta = 0.1$}
        \label{fig:plot2b}
    \end{subfigure}
    \begin{subfigure}[b]{0.32\textwidth}
        \centering
        \includegraphics[width=\textwidth]{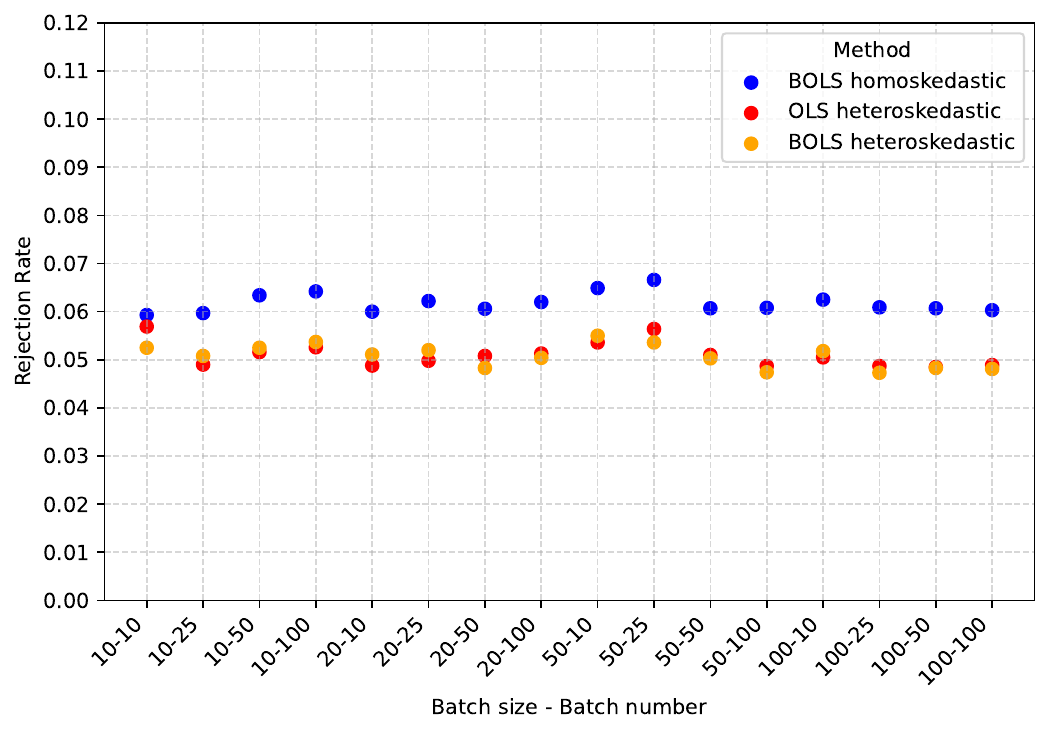}
        \caption{$\Delta = 0.2$}
        \label{fig:plot2c}
    \end{subfigure}
    \scriptsize
     \begin{justify}
       \textit{Notes:} This figure shows Monte Carlo simulation results for the Bernoulli Thompson Sampling algorithm. Combinations of batch size and number of batches increase along the horizontal axis. Rejection rates are denoted on the vertical axis. The parameter $\Delta$ indicates the difference between the true expected values of both arms. Each circle shows the average rejection rate for the given test statistic which is indicated by color. For each batch size / number of batches combinations 10,000 repetitions were executed. In panel (a) the success probability for arm $1$ is $p_1 = 0.5$ and arm $2$ is also $p_2 = 0.5$. In panel (b) it is $p_1 = 0.6$ and $p_2 = 0.5$. For panel (c) it is  $p_1 = 0.7$ and $p_2 = 0.5$.
          \end{justify}
    \caption{Bernoulli Thompson Sampling}
    \label{fig:sim3}
\end{figure}

\paragraph{Simulation II: Rejection Rates at Small and Large Margins in Small and Large Samples}

To study behavior in smaller samples, we vary the number of batches (10--100) and batch sizes (10--100). Each configuration is repeated 10{,}000 times. Figures~\ref{fig:sim2} and~\ref{fig:sim3} report rejection rates of a 5\% significance level test of the true null $H_0:\Delta=\delta_0$.

For $\varepsilon$-Greedy (Figure~\ref{fig:sim2}), the homoskedastic BOLS statistic is unreliable in all settings: it overrejects sharply at $\Delta=0$ and underrejects for positive margins. The heteroskedasticity-robust OLS statistic performs reasonably well and improves as the margin grows. Across all designs, our heteroskedasticity-robust BOLS statistic maintains rejection rates close to 5\%.

For Thompson Sampling (Figure~\ref{fig:sim3}), the zero-margin case exhibits no heteroskedasticity, so the heteroskedastic and the homoskedastic BOLS statistics perform well with rejection rates near 5\%. The heteroskedasticity-robust OLS overrejects somewhat. When margins increase, inducing heteroskedasticity, the homoskedastic BOLS statistic begins to overreject, while both robust statistics remain close to nominal size.

\section{Conclusion}

Adaptive experiments have made precision-equalizing inference increasingly relevant in sequential settings where treatment assignment depends on past outcomes. The BOLS inverse-standard-error-weighted statistic provides a simple, asymptotically valid procedure for inference under heteroskedasticity, extending previous results derived for the homoskedastic case.

\bibliographystyle{econometrica_abbrv}
\bibliography{references}

\newpage
\appendix
\section{Weight structure under homoskedasticity and heteroskedasticity}

\begingroup
\renewcommand{\arraystretch}{1.8}
\begin{table}[htbp]
\caption{Weight structure under homoskedasticity and heteroskedasticity for two-armed experiments\label{tab:weights}}
\centering
\begin{tabular}{@{}lcc@{}}
\toprule
 & \textbf{Non-adaptive} & \textbf{Adaptive} \\
\midrule
\multicolumn{3}{c}{\textit{Homoskedastic case: $\sigma_{1,t}^2 = \sigma_{0,t}^2 = \sigma^2$}}\\
Realized share $\hat\pi_t$
& $\hat\pi_t = \pi_t$, constant across batches
& $\pi_t$ depends on $\mathcal{H}_{t-1}$, $\hat\pi_t$ random \\
Variance $v_t$
& $\displaystyle v_t = \frac{\sigma^2}{n_t\,\pi_t(1-\pi_t)}$
& $\displaystyle v_t = \frac{\sigma^2}{n_t\,\hat\pi_t(1-\hat\pi_t)}$ \\
Weight $\displaystyle \omega_t =\frac{w_t}{S}$
& $\displaystyle \frac{\sqrt{n_t}}{\sum_{s=1}^T \sqrt{n_s}}$
& $\displaystyle \frac{\sqrt{n_t\,\hat\pi_t(1-\hat\pi_t)}}{\sum_{s=1}^T \sqrt{n_s\,\hat\pi_s(1-\hat\pi_s)}}$ \\
\midrule
\multicolumn{3}{c}{\textit{Heteroskedastic case: $\sigma_{1,t}^2\neq\sigma_{0,t}^2$}}\\
Realized share $\hat\pi_t$
& $\hat\pi_t = \pi_t$, constant across batches
& $\pi_t$ depends on $\mathcal{H}_{t-1}$, $\hat\pi_t$ random \\
Variance $v_t$
& $\displaystyle v_t = \frac{1}{n_t}\!\left(\frac{\sigma_{1,t}^2}{\pi_t}+\frac{\sigma_{0,t}^2}{1-\pi_t}\right)$
& $\displaystyle v_t = \frac{1}{n_t}\!\left(\frac{\sigma_{1,t}^2}{\hat\pi_t}+\frac{\sigma_{0,t}^2}{1-\hat\pi_t}\right)$ \\
Weight $\displaystyle \omega_t=\frac{w_t}{S}$
& $\displaystyle \frac{\sqrt{n_t}\!\left(\frac{\sigma_{1,t}^2}{\pi_t}+\frac{\sigma_{0,t}^2}{1-\pi_t}\right)^{-1/2}}{\sum_{s=1}^T \sqrt{n_s}\!\left(\frac{\sigma_{1,s}^2}{\pi_s}+\frac{\sigma_{0,s}^2}{1-\pi_s}\right)^{-1/2}}$
& $\displaystyle \frac{\sqrt{n_t}\!\left(\frac{\sigma_{1,t}^2}{\hat\pi_t}+\frac{\sigma_{0,t}^2}{1-\hat\pi_t}\right)^{-1/2}}{\sum_{s=1}^T \sqrt{n_s}\!\left(\frac{\sigma_{1,s}^2}{\hat\pi_s}+\frac{\sigma_{0,s}^2}{1-\hat\pi_s}\right)^{-1/2}}$ \\
\bottomrule
\end{tabular}

\vspace{0.5em}
\raggedright
\footnotesize
\textit{Notes:} $n_t = N_{1,t} + N_{0,t}$ is total batch size, and $\hat\pi_t = N_{1,t}/n_t$ is the realized treatment share. The corresponding algorithmic assignment probability is denoted by $\pi_t$. The non-adaptive column refers to fixed-count designs, in which $N_{1,t}=\pi_t n_t$ is set in advance (with $\pi_t n_t$ an integer) and $\pi_t$ is the same in every batch, so that $\hat\pi_t=\pi_t$ exactly. In the adaptive case $\pi_t$ is $\mathcal{H}_{t-1}$-measurable and the weights depend on the realized shares. Inverse-standard-error weights equalize the predictable variance of the batchwise contributions and are $w_t = 1/\sqrt{v_t}$, their sum $S = \sum_{s=1}^T w_s$. Under homoskedasticity, adaptive weights $\omega_t$ depend on both sample size and realized balance $\hat\pi_t(1-\hat\pi_t)$, whereas in the non-adaptive case $\pi_t(1-\pi_t)$ is common to all batches and cancels, leaving weights that depend on $n_t$ alone. Under heteroskedasticity, weights additionally adjust for treatment-specific outcome variances $\sigma_{a,t}^2$.
\end{table}
\endgroup

\section{Proof of Theorem \ref{thm:main}}
\label{app:proof}

We prove asymptotic normality of $Z_{\mathrm{het}}$ using a triangular-array martingale central limit theorem applied directly to the individual-level observations across all batches. Throughout, $T$ is fixed and $n=\min_t n_t\to\infty$. All quantities may depend on $n$; this dependence is suppressed for readability.
\subsection*{B.1 Setup}

We first condition on the realized assignment pattern in each batch. Recall that
\(H_{t-1}\) denotes the history before batch \(t\), and define
\begin{equation}
    \mathcal G_t
    =
    H_{t-1}\vee\sigma(A_{1,t},\ldots,A_{n_t,t}).
\end{equation}
Thus \(\mathcal G_t\) contains the pre-batch history and the full assignment
vector in batch \(t\), but not the batch-\(t\) outcomes. Conditional on
\(\mathcal G_t\), the realized treatment counts
\begin{equation}
    N_{1,t}=\sum_{i=1}^{n_t}A_{i,t},
    \qquad
    N_{0,t}=\sum_{i=1}^{n_t}(1-A_{i,t})
\end{equation}
are known. The variance parameters \(\sigma_{a,t}^2\) are fixed arm- and
batch-specific quantities satisfying, by Assumption~\ref{ass:moments},
\begin{equation}
    E[\varepsilon_{i,t}(a)^2\mid H_{t-1}]
    =
    \sigma_{a,t}^2 .
\end{equation}
Hence
\begin{equation}
    v_t
    =
    \frac{\sigma_{1,t}^2}{N_{1,t}}
    +
    \frac{\sigma_{0,t}^2}{N_{0,t}}
\end{equation}
is \(\mathcal G_t\)-measurable and can be treated as fixed in conditional
variance calculations.

By the potential-outcome representation in equation~\eqref{eq:potentialoutcomes},
the observed outcome is
\begin{equation}
Y_{i,t}
=
A_{i,t}Y_{i,t}(1)+(1-A_{i,t})Y_{i,t}(0)
=
A_{i,t}\{\mu_{1,t}+\varepsilon_{i,t}(1)\}
+
(1-A_{i,t})\{\mu_{0,t}+\varepsilon_{i,t}(0)\}.
\end{equation}

Substituting this into the means of equation~\eqref{eq:samplemeans} gives
\begin{equation}
\frac{1}{N_{1,t}}\sum_{i=1}^{n_t}A_{i,t}Y_{i,t}
=
\frac{1}{N_{1,t}}\sum_{i=1}^{n_t}
A_{i,t}\{\mu_{1,t}+\varepsilon_{i,t}(1)\}
=
\mu_{1,t}
+
\frac{1}{N_{1,t}}\sum_{i=1}^{n_t}
A_{i,t}\varepsilon_{i,t}(1),
\end{equation}
because \(\sum_{i=1}^{n_t}A_{i,t}=N_{1,t}\). Similarly,
\begin{equation}
\frac{1}{N_{0,t}}\sum_{i=1}^{n_t}(1-A_{i,t})Y_{i,t}
=
\mu_{0,t}
+
\frac{1}{N_{0,t}}\sum_{i=1}^{n_t}
(1-A_{i,t})\varepsilon_{i,t}(0).
\end{equation}
Hence
\begin{equation}
\widehat\Delta_t
=
\mu_{1,t}-\mu_{0,t}
+
\frac{1}{N_{1,t}}\sum_{i=1}^{n_t}A_{i,t}\varepsilon_{i,t}(1)
-
\frac{1}{N_{0,t}}\sum_{i=1}^{n_t}(1-A_{i,t})\varepsilon_{i,t}(0).
\end{equation}
Since \(\Delta_t=\mu_{1,t}-\mu_{0,t}\), subtracting \(\Delta_t\) yields
\begin{equation}
\widehat\Delta_t-\Delta_t
=
\frac{1}{N_{1,t}}\sum_{i=1}^{n_t}A_{i,t}\varepsilon_{i,t}(1)
-
\frac{1}{N_{0,t}}\sum_{i=1}^{n_t}(1-A_{i,t})\varepsilon_{i,t}(0).
\end{equation}

\subsection*{B.2 Martingale difference representation}

For each batch \(t\), define the within-batch filtration
\begin{equation}
    \mathcal F_{t,0}=\mathcal G_t,
    \qquad
    \mathcal F_{t,i}
    =
    \mathcal G_t\vee\sigma(Y_{1,t},\ldots,Y_{i,t}),
    \quad i=1,\ldots,n_t.
\end{equation}
The filtrations are concatenated across batches in lexicographic order. Thus \(\mathcal F_{t,i-1}\) contains the history before batch \(t\), the full assignment vector in batch \(t\), and the first \(i-1\) outcomes \(Y_{1,t},\ldots,Y_{i-1,t}\), but not the current outcome \(Y_{i,t}\). Therefore the current centered shock still has conditional mean zero.

Define the realized centered outcome shock
\begin{equation}
    \widetilde\varepsilon_{i,t}
    =
    (1-A_{i,t})\varepsilon_{i,t}(0)
    +
    A_{i,t}\varepsilon_{i,t}(1).
\end{equation}
Because \(A_{i,t}\in\{0,1\}\), we have \(A_{i,t}^2=A_{i,t}\), \((1-A_{i,t})^2=1-A_{i,t}\), and \(A_{i,t}(1-A_{i,t})=0\), so the batchwise estimation error can equivalently be expressed as
\begin{equation}
    \widehat\Delta_t-\Delta_t
    =
    \sum_{i=1}^{n_t}
    \left(
        \frac{A_{i,t}}{N_{1,t}}
        -
        \frac{1-A_{i,t}}{N_{0,t}}
    \right)
    \widetilde\varepsilon_{i,t}.
\end{equation}

The statistic \(Z_{\mathrm{het}}\) is the \(1/\sqrt T\)-normalized sum of the batchwise
studentized estimation errors. We therefore define the individual contribution
\begin{equation}
    D_{t,i}
    =
    \frac{1}{\sqrt T}\frac{1}{\sqrt{v_t}}
    \left(
        \frac{A_{i,t}}{N_{1,t}}
        -
        \frac{1-A_{i,t}}{N_{0,t}}
    \right)
    \widetilde\varepsilon_{i,t}.
\end{equation}
Since \(\Delta_t=\delta_{0,t}\) under the batchwise null,
\begin{equation}
Z_{\mathrm{het}}
=
\frac{1}{\sqrt T}
\sum_{t=1}^T
\frac{\widehat\Delta_t-\Delta_t}{\sqrt{v_t}}
=
\sum_{t=1}^T\sum_{i=1}^{n_t}D_{t,i}.
\end{equation}

Since \(A_{i,t}\), \(N_{1,t}\), \(N_{0,t}\), and \(v_t\) are
\(\mathcal F_{t,i-1}\)-measurable, only the centered outcome shock
\(\widetilde\varepsilon_{i,t}\) remains random at step \(i\). By
Assumption~\ref{ass:indep}, the potential-outcome shocks are independent of
the batch-\(t\) randomization and independent across individuals within the
batch, conditional on \(H_{t-1}\). By Assumption~\ref{ass:moments}, these
shocks have conditional mean zero:
\begin{equation}
    E[\widetilde\varepsilon_{i,t}\mid \mathcal F_{t,i-1}]
    =
    0.
\end{equation}
Therefore,
\begin{equation}
    E[D_{t,i}\mid \mathcal F_{t,i-1}]
    =
    0,
\end{equation}
so \(\{D_{t,i}\}\) is a triangular array of martingale differences.

\subsection*{B.3 Predictable quadratic variation}

The conditional second moment of $D_{t,i}$ is
\begin{equation}
    E[D_{t,i}^2\mid\mathcal{F}_{t,i-1}] = \frac{1}{T}\frac{1}{v_t}\left[\frac{A_{i,t}\,\sigma_{1,t}^2}{N_{1,t}^2} + \frac{(1-A_{i,t})\,\sigma_{0,t}^2}{N_{0,t}^2}\right].
\end{equation}
Summing over $i$ within batch $t$:
\begin{equation}
    \sum_{i=1}^{n_t} E[D_{t,i}^2\mid\mathcal{F}_{t,i-1}] = \frac{1}{T}\frac{1}{v_t}\left[\frac{\sigma_{1,t}^2}{N_{1,t}} + \frac{\sigma_{0,t}^2}{N_{0,t}}\right] = \frac{1}{T}.
\end{equation}
Therefore,
\begin{equation}
    \sum_{t=1}^T\sum_{i=1}^{n_t} E[D_{t,i}^2\mid\mathcal{F}_{t,i-1}] = 1.
\end{equation}
The predictable quadratic variation is exactly normalized to one. Because of the local studentization no random quantity remains in the variance accumulation.

\subsection*{B.4 Lindeberg condition}

To rule out large individual observations, the clipping condition is relevant. It ensures that both treatment arms receive a number of observations proportional to \(n_t\), so no single observation receives too much weight. By Assumption~\ref{ass:clip} and the conditional law of large numbers, with probability approaching one there exists \(\underline\kappa>0\) such that $N_{1,t}\geq \underline\kappa n_t, \ N_{0,t}\geq \underline\kappa n_t$ for all \(t\). On this event, together with the uniform variance bounds in Assumption~\ref{ass:moments}, $
v_t
=
\frac{\sigma_{1,t}^2}{N_{1,t}}
+
\frac{\sigma_{0,t}^2}{N_{0,t}}
\asymp n_t^{-1}.
$ This means that \(v_t\) is bounded above and below by constant multiples of \(1/n_t\).
While \(N_{1,t}^{-1}\) and \(N_{0,t}^{-1}\) are both of order \(n_t^{-1}\), \(v_t^{-1/2}\asymp n_t^{1/2}\). Therefore each coefficient multiplying a centered outcome in \(D_{t,i}\) is of order \(n_t^{-1/2}\).

Using the bounded \((2+\delta)\)-th moment condition in Assumption~\ref{ass:moments}, there exists \(C<\infty\) such that
\begin{equation}
    \sum_{t=1}^T\sum_{i=1}^{n_t}
    E\!\left[
    |D_{t,i}|^{2+\delta}
    \mid\mathcal{F}_{t,i-1}
    \right]
    \leq
    C\sum_{t=1}^T n_t^{-\delta/2}
    \longrightarrow 0.
\end{equation}
This Lyapunov condition rules out large individual contributions and, by Markov's inequality, implies the martingale Lindeberg condition: for every
\(\varepsilon>0\),
\begin{equation}
    \sum_{t=1}^T\sum_{i=1}^{n_t}
    E\!\left[
    D_{t,i}^2\,\ind\{|D_{t,i}|>\varepsilon\}
    \mid\mathcal{F}_{t,i-1}
    \right]
    \plim 0.
\end{equation}

The martingale Lindeberg condition rules out asymptotically relevant outliers.

\subsection*{B.5 Martingale central limit theorem}

The triangular-array martingale difference sequence satisfies: (i) the predictable quadratic variation equals $1$ exactly, and (ii) the Lindeberg condition holds. By the triangular-array martingale CLT \citep[Theorem~3.2]{Hall1980},
\begin{equation}
    Z_{\mathrm{het}} = \sum_{t=1}^T\sum_{i=1}^{n_t} D_{t,i} \;\overset{d}{\longrightarrow}\; \mathcal{N}(0,1).
\end{equation}
This proves the infeasible part of Theorem~\ref{thm:main}. \qed

\paragraph{Remark (Stacked scores and other projections)}
The same argument also gives asymptotic normality of the stacked batchwise test statistic
\begin{equation}
\mathbf z_{\mathrm{het}}
=
\left(
\frac{\widehat\Delta_1-\Delta_1}{\sqrt{v_1}},
\ldots,
\frac{\widehat\Delta_T-\Delta_T}{\sqrt{v_T}}
\right)'
\overset{d}{\longrightarrow}
N(0,I_T).
\end{equation}
Indeed, replacing the scalar weight \(1/\sqrt T\) in \(D_{t,i}\) by any fixed
coefficient \(c_t\) gives
\begin{equation}
c'\mathbf z_{\mathrm{het}}
=
\sum_{t=1}^T
c_t\frac{\widehat\Delta_t-\Delta_t}{\sqrt{v_t}}
\overset{d}{\longrightarrow}
N(0,c'c),
\end{equation}
so the claim follows by Cramér--Wold. The statistic
\(Z_{\mathrm{het}}\) is the special case
\(c=T^{-1/2}\mathbf 1_T\) under the batchwise null
\(\Delta_t=\delta_{0,t}\) for all \(t\). For a common treatment effect,
\(\Delta_t=\Delta\), this is the natural projection. Other pre-specified
projections are mainly useful for testing time variation in \(\Delta_t\), such
as linear trends or early-versus-late contrasts.
\subsection*{B.6 Feasible statistic}

By Assumption~\ref{ass:var}, $\hat\sigma_{a,t}^2\plim\sigma_{a,t}^2$. Since $T$ is fixed and, with probability approaching one, the treatment counts are bounded away from zero at rate $n_t$, it follows that $\max_{1\leq t\leq T}|\hat v_t/v_t - 1|\plim 0$. Since
\begin{equation}
    \frac{\hat\Delta_t-\Delta_t}{\sqrt{\hat v_t}} = \frac{\hat\Delta_t-\Delta_t}{\sqrt{v_t}}\left(\frac{v_t}{\hat v_t}\right)^{1/2},
\end{equation}
the infeasible batch scores are tight and the multiplicative factor converges to one uniformly over fixed $T$. Hence $\hat Z_{\mathrm{het}} - Z_{\mathrm{het}} = o_p(1)$, and by Slutsky's theorem, $\hat Z_{\mathrm{het}}\overset{d}{\longrightarrow}\mathcal{N}(0,1)$. \qed

\end{document}